\documentclass[aps,prb,a4paper,twocolumn,showpacs,floatfix,citeautoscript]{revtex4}
\usepackage{graphicx}
\usepackage{amsfonts}
\usepackage{amsmath}
\usepackage{amssymb}
\usepackage{bm}
\usepackage{dcolumn}
\usepackage{hyperref}
\usepackage[usenames]{color}
\hypersetup{pdfborder=0 0 0,colorlinks=true,citecolor=blue,linkcolor=blue}
\begin{document}

\title{Quantum confinement in perovskite oxide heterostructures:\\
tight binding instead of nearly free electron picture}
\author{Zhicheng Zhong$^1$, Qinfang Zhang$^2$, and Karsten Held$^1$}
\affiliation{$^1$Institute of Solid State Physics, Vienna University of Technology, A-1040 Vienna, Austria \\
$^2$Key Laboratory for Advanced Technology in Environmental Protection of Jiangsu Province,
Yancheng Institute of technology, China}
\date{\today}

\begin{abstract}
Most recently, orbital-selective quantum well states of $d$ electrons have been experimentally observed in SrVO$_3$ ultrathin films [K. Yoshimatsu et. al., Science 333, 319 (2011)] and SrTiO$_3$ surfaces [A. F. Santander-Syro et. al., Nature 469, 189 (2011)]. Hitherto, one tries to explain these experiments by a nearly free electron (NFE) model, an approach widely used for delocalized electrons in semiconductor heterostructures and simple metal films. We show that a tight binding (TB) model is more suitable for describing heterostructures with more localized $d$ electrons. In this paper, we construct from first principles simple TB models for perovskite oxide heterostructures and surfaces. We show that the TB model provides  a simple intuitive physical picture and yields, already with only two parameters, quantitatively much more reliable results, consistent with experiment. 
\end{abstract}
\maketitle

\section{Introduction}
In bulk transition metal oxides, $d$ electrons exhibit interesting and intriguing electronic and magnetic properties.\cite{Imada:rmp98} Thanks to recent progress of epitaxial growth techniques, perovskite oxide heterostructures can now be made and controlled at atomic scales so that $d$ electrons are confined within a region of a few unit cells ($\sim$1$nm$) in the $z$-direction of the epitaxial growth\cite{Zubko,Mannhart:sci10,Huijben:adv09}. As a result of the confinement, many novel physical phenomena occur, including orbital-selective quantum well states\cite{Yoshimatsu:sci11,Santander:nat11,zhong:epl12}, metal-insulator transitions and superconductivity tunable by gate voltage,\cite{Ohtomo:nat04,Thiel:sc06,Caviglia:nat08} enhanced thermoelectric effects,\cite{OHTA:natm07} thickness dependent ferromagnetism,\cite{Chang:prl09,Verissimo:prl12,Luder:prb09} strong spin-orbit coupling effects \cite{Caviglia:prl10b,Shalom:prl10b,Zhong:prb13,Khalsa:arxiv13,Joshua:nc12}, tunable correlation \cite{Moetakef:prb12, Monkman:natm12} and a rich variety of phases including spin, charge and orbital ordering \cite{Okamoto:nat04}.

A direct consequence of the confinement at the interface are, in particular, quantum well states, which can serve as a starting point for other complex physical phenomena. Very recently, quantum well states have been convincingly observed experimentally by means of angle-resolved photoemission spectroscopy for two distinct oxide heterostructures: (i) SrVO$_3$ (SVO) ultrathin films \cite{Yoshimatsu:sci11}, where electrons are geometrically confined inside the film, and (ii) SrTiO$_{3}$ (STO) surfaces\cite{Santander:nat11} (which can be considered as a STO/vacuum heterostructure), where electrons are confined in a two-dimension (2D) surface potential well \cite{Mannhart:sci10}. In both cases, very similar orbital-selective quantum well states are observed: $d$ electrons with $yz$/$xz$ orbital characters exhibit a large quantization of the energy levels, whereas $xy$ electrons exhibit a much smaller level spacing. This behavior has been ascribed to a nearly free electron (NFE) model in the literature\cite{Yoshimatsu:sci11,Santander:nat11}.

The NFE model is widely used in semiconductors heterostructures as well as for simple metal thin films\cite{Hicks:prb93,Milun:rpp02}. Electrons are regarded to move almost freely with an energy vs.\ momentum ($\vec{k}$) dispersion relation $\frac{\hbar ^{2} k^{2}}{2m^*}$ in terms of the effective mass $m^*$. The confinement in the $z$ direction is described by a potential well $V(z)$ of a characteristic length of 10$nm$. Such a simple model with only two variables $m^*$ and $V(z)$ works perfectly for semiconductor heterostructures. However, its applicability to oxide heterostructure is questionable, because it is well known that $d$ electrons are much more localized than the $s,p$ electrons in semiconductor heterostructures. In a perovskite oxide, an electron is tightly bound to a transition metal ion site and moves in the crystal structure by hopping from one site to a neighboring site. One  might therefore expect that a tight binding (TB) model will give a much better description of oxide heterostructure than the NFE model. While finalizing this paper, a related work by Park and Millis \cite{Park:arxiv13} suggested a NFE model in-plane and a tight-binding model out of plane, where no hopping terms along  $\vec{R}=(1,0,1)$ and $\vec{R}=(1,1,0)$ are considered and $yz$ orbital in the $\Gamma-X$ direction is dispersionless. Hence, while the TB modeling appears natural and first steps have been undertaken in this direction \cite{Popovic:prl08,Chang:prl09,Stengel:prl11,Verissimo:prl12}, a systematic comparison between TB and NFE model for oxide heterostructures is hitherto missing. Similarly, there has not been a systematic investigation of how many TB parameters are needed for an accurate description.
Hence, it is unclear at present how complicated or simple the TB 
description actually is for such heterostructures.

In this paper, we do first-principles density functional theory (DFT) calculations and construct from these TB models for describing the quantum well states in perovskite oxide heterostructures and surfaces. We further simplify our models to an effective hopping term $t$ and a local potential term $\varepsilon$, instead of $m^*$ and $V(z)$ for the NFE model. We show that for thin SVO films, the geometrical confinement is described by cutting the hopping term from surface layer to vacuum. The quantized energies are $2t \cos(\frac{\pi n}{N+1})$, where N is the thickness of the film and $n$ is a quantum number, ranging from 1 to $N$. In contrast, the NFE model yields
 $\frac{{\hbar}^2\pi^{2}n^2}{2m^{*}N^2a^{2}}$ where $a$ is the lattice constant of bulk SVO. Moreover, we study the potential well confinement at STO surfaces or LaAlO$_{3}$/SrTiO$_3$ (LAO/STO) heterostructures. 
Here, we need to include a layer-$i$ dependent potential $\varepsilon_{i}$ in our model. For a realistic potential well, we find the lowest quantized $yz$/$xz$ state is on the verge of becoming a surface bound state. Hence, its spatial distribution can be easily tuned by a gate voltage or an electric field. Our results show that the TB approach, instead of the NFE approach, is the natural basis for modelling heterostructures of transition metal oxides. Our TB model can serve as a starting point for follow-up studies such as advanced transport or many-body effects.

\section{DFT study of SVO bulk and thin films}
\subsection{Bulk SVO}
\begin{figure}[t!]
\includegraphics[width=\columnwidth]{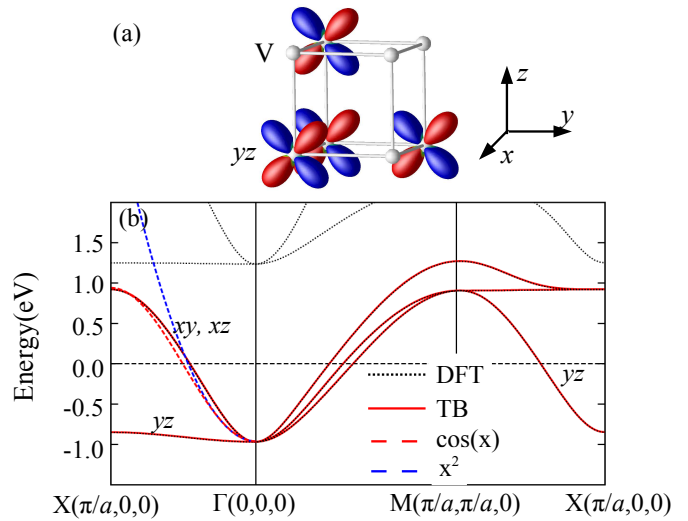}
\caption{(a) Schematic figure of V sites in bulk SVO and V 3$d$ $yz$ orbitals with lobes expanding in the $y$-$z$ plane. (b) SVO band structure calculated by DFT (black dotted lines) and compared to the $t_{2g}$ TB Hamiltonian Eq.(\ref{H})
(red solid lines). Along X-$\Gamma$, a red dashed line indicates a fit to
a cosine function, and the blue one a fit to a parabolic function.}
\label{Fig1}
\end{figure}

Bulk SVO (Fig.\ref{Fig1}) is a non-magnetic correlated metal with perfect cubic perovskite structure of space symmetry group 221 $Pm$-$3m$. When studying its thin film growth along the (001) direction, we usually regard it as an alternating stacking of SrO and VO$_2$ layers. In this paper, we study symmetric SVO thin films containing $N$ layers of VO$_2$ and $N+1$ layers of SrO so that the surfaces are SrO terminated, see Fig.\ref{Fig2}. We employ a sufficiently thick vacuum of 10\AA\ for the supercell calculation and vary  the thickness $N$ one to ten SVO unit cells. We fix the in-plane lattice constant to the calculated equilibrium bulk value $a_{\rm{SVO}}$=3.86\AA, and optimize the internal coordinates.  Our DFT results reveal that the surface oxygen atom  relax outward by 0.06\AA, while the Sr atom relax inward by 0.12\AA; the relaxation of other atoms is negligible. We note that including STO as a substrate or making a SVO/STO superlattice will not change our main conclusion. A VO$_2$ terminated surface instead of a SrO one
on the other hand is rather different as this breaks the VO$_6$ octahedral crystal field of bulk SVO. 

First-principles density-functional-theory (DFT) calculations are performed using the all-electron full-potential augmented plane-wave method in the Wien2k\cite{WIEN2k} implementation. We use the generalized gradient approximation (GGA)\cite{PerdewPRL96} of the exchange-correlation potential and 10$\times$10$\times$1 $k$-point grid.
 Let us note in passing that including a on-site Coulomb interaction U within the DFT+U method does not improve the calculations: It cannot give renormalized bands; and, even worse, it will give a magnetic ground state inconsistent with experimental observations.

Considering the formal charge valence Sr$^{2+}$, V$^{4+}$ and O$^{2-}$, bulk SVO has a $d^{1}$ electronic configuration with one electron in the V 3$d$ states. Due to the crystal field splitting of the VO$_{6}$ octahedron, three $t_{2g}$ states ($xy$, $yz$, $xz$)  are shifted down in energy while two $e_g$ states are pushed up. Thus, one electron will partially fill three $t_{2g}$ orbitals centered at the V sites. One of the $t_{2g}$ orbitals, the $yz$ orbital, is schematically shown in Fig.\ref{Fig1}(a) as an example. This orbital predominantly 
expands in the $y$-$z$ plane, and a pair of its lobes point
to a corresponding pair of lobes from $yz$ orbital at nearest neighbor sites in the $y$ and $z$ direction, see Fig.\ref{Fig1}(a). The other two orbitals, i.e, $xy$ and $xz$, have the same character and are related by cubic symmetry 
 to 
the $yz$ orbital (i.e., $z\leftrightarrow x$ and $y\leftrightarrow x$, respectively).

Our DFT calculations show: Below the Fermi energy, O$_{2p}$ states are located between -7.2eV and -2.1eV; near the Fermi energy, Ti 3$d$ $t_{2g}$ bands are found
between -0.96eV and 1.47 eV and are  slightly hybridized with O$_{2p}$; above the Fermi energy, $e_g$ states are located between 1.2eV to 5eV. Thus, as expected, the O$_{2p}$ states are fully occupied, V $e_g$ states are empty, and three $t_{2g}$ states are partially filled with one electron in total, equally distributed to the three orbitals because of the cubic symmetry. The $t_{2g}$ bands are plotted in Fig.\ref{Fig1}(b) along high symmetric $k$ lines. The total bandwidth is 2.5eV, ranging from -0.96eV at $\Gamma (0,0,0)$ to 1.47eV at R$(\pi/a,\pi/a,\pi/a)$. At $\Gamma$ the three bands are degenerate. Along $\Gamma$-X($\pi/a$,0,0), the $yz$ band has a much smaller energy dispersion of only 0.12eV, whereas the two $xy$/$xz$ are degenerate in this direction and have a much larger energy dispersion of 1.9eV.

Around $\Gamma$, we fit the DFT bands by a parabolic energy dispersion of nearly free electron $\frac{\hbar ^{2} k^{2}}{2m^*}$ as shown in Fig.\ref{Fig1}(b), and obtain effective masses $m^*$=0.56$m_e$ for the $xy$/$xz$ bands and $m^*$=8.4$m_e$ for the $yz$ band, where $m_e$ is free electron mass. Considering the rotation symmetry of the $t_{2g}$ orbitals, we can argue that carriers with $yz$ characters are light in the $y$ and $z$ direction, but heavy in the $x$ direction. In other words, along a specific direction such as $z$, there are two light carriers ($yz$ and $xz$), and one heavy carrier ($xy$) \cite{Footnoot1}. At higher energies, e.g. towards the X point, the energy dispersion of the $t_{2g}$ bands follows a cosine function, instead of a parabolic function of NFE, see Fig.\ref{Fig1}(b). Since this high-energy part does contribute to the quantized energies in heterostructures, the applicability of the NFE model for describing SVO thin films becomes questionable.

\subsection{SVO thin films}
The unit cell of SVO thin films have $N$ layers of Vanadium, containing 3$N$ $t_{2g}$ orbitals in the supercell. In our notation, the $z$ axis denotes the out of plane direction and the $x$ and $y$ axis the in plane directions. An electron is allowed to move only in-plane whereas it is confined by the film in the $z$ direction.
Hence, instead of a dispersion in the $z$ direction, we obtain $N$ quantized levels for each orbital and in-plane $k$ point. 

\begin{figure}[t!]
\includegraphics[width=\columnwidth]{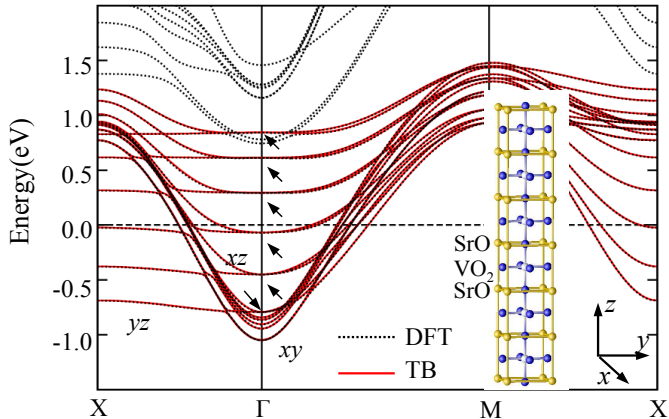}
\caption{Band structure of a six layers SVO thin film calculated by DFT (black dotted lines) and the TB Hamiltonian (red solid lines). The confinement in the $z$-direction leads to quantized energies levels which are indicated by arrows 
for the $yz$/$xz$ orbitals at $\Gamma$. Inset: atomic structure of the SVO thin film.}
\label{Fig2}
\end{figure}

The DFT calculated band structure for $N$=6 layers is plotted in Fig.\ref{Fig2} along high-symmetric in-plane $k$ points. In total, 3$\times$6 $t_{2g}$ bands are located between -1.0 to 1.5eV. Analyzing the symmetry of the bands as well as projecting on each orbital and site, we are able to identify the character of all bands. When going from SVO bulk to thin films, the  translation symmetry along the $z$ direction is broken, whereas the in-plane translational and rotational symmetry remains. Therefore, the initial triply degenerate states at $\Gamma$ split into a $xy$ state and a doubly degenerate $yz/xz$ state.

At $\Gamma$, the lowest band is of purely surface V $xy$ character, followed by the $xy$ orbitals of the second and third layer. The surface $xy$ band is 0.16eV lower than the other $xy$ bands which are all close in energy. Such a band splitting arises from a local potential drop of the surface layer, as revealed by the Wannier projection discussed below in Table \ref{interface}. Here, we see that the dispersion of all $xy$ bands is similar to that of bulk. This is because $xy$ orbitals expand mainly in-plane, and the confinement along the $z$ direction has hence little influence.

Turning to the $yz/xz$ orbitals, we note that
 $yz$ has a small and $xz$ a large energy-momentum dispersion 
along $\Gamma$-X, i.e., in the $x$ direction. Of course the behavior is opposite in the $y$ direction, and the two orbitals are degenerate at $\Gamma$. In contrast to the $xy$ bands, the two $yz/xz$ orbitals exhibit a pronounced energy subband structure: Six discrete energies are separated by an energy level spacing of about 300meV. This is because the $yz$/$xz$ orbitals expand in the $z$ direction. Along this direction, their energy dispersion is large and hence the confinement along $z$ leads to a pronounced energy quantization if the electrons are confined in a thin film. Projecting the $yz$/$xz$ states onto each site (not shown) reveals that all quantized $yz$/$xz$ states do not belong to a single layer, but indeed spread throughout the thin film. Hence in contrast to the $xy$ bands, each $yz$ subband is a superposition of $yz$ orbitals from all layers.

\begin{figure}[t!]
\includegraphics[width=\columnwidth]{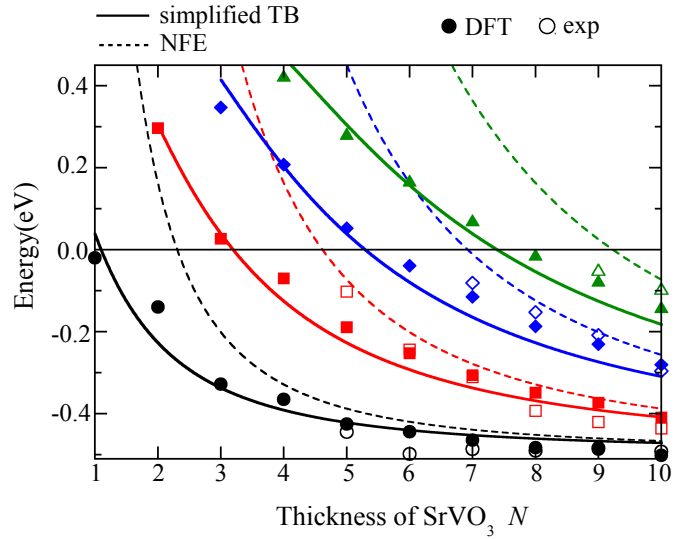}
\caption{Quantized energies of the quantum well states of the V $yz$ orbitals as a function of SVO film thickness $N$ at the $\Gamma$ point. Results with quantum numbers $n$=1-4 are shown in black, red, blue and green, respectively. Experimental results (unfilled symbols) are taken from Ref.\ \onlinecite{Yoshimatsu:sci11}; DFT results with a renormalization factor $Z=1.8$ (filled symbols) are extracted from the band structures of SVO thin films with different thickness $N$, in the same way as shown in Fig.\ref{Fig2} for $N$=6. The NFE models gives $\varepsilon+2t+\frac{{\hbar}^2\pi^{2}n^2}{2m^{*}N^2a^{2}}$ (dashed lines),
and the TB model $\varepsilon+2t\cos(\frac{\pi n}{N+1})$ (Eq.(\ref{quantized}) solid lines). Here, $m^{*}=-\frac{\hbar ^{2}}{2a^2t}=0.53m_{e}$ and respectively $t=-0.475$eV, $\varepsilon=-0.01eV$ are estimated from the DFT, which yields consistency results in the limit of $N \gg n$.}
\label{Fig3}
\end{figure}

The subband energy structure of the $yz$ orbitals at $\Gamma$ (arrows in Fig.\ref{Fig2}) have been experimentally observed in angular-resolved photoemission spectroscopy (ARPES) \cite{Yoshimatsu:sci11}.
Fig.\ref{Fig3} shows the comparison of experiment 
and theory for varying film thickness $N$.
To account for correlation effects beyond DFT, we have renormalized the DFT band structure by a factor of 1/$Z$ with a renormalization factor $Z$=1.8
taken from bulk SVO \cite{Yoshida:PRL05,Nekrasov:prb06,Pavarini:njp05}. Clearly, there is a good agreement of
theory and experiment regarding the magnitude and the general behavior of the quantized energy levels. We note that a metal-to-insulator transition occurs for SVO ultrathin film with $N \leq$2 \cite{Yoshimatsu:prl10}, and hence the picture of renormalized quasi-particle fails in that region.

Considering the good agreement between DFT and experimental results, we now try to extract a simple model based on the DFT results, for describing the quantum confinement. In a NFE model, the geometrical confinement of SVO thin films is approximated by an infinite potential well, where the wavefunction at the boundary is hence zero. Such a boundary condition results in quantized energy levels with energies $\frac{{\hbar}^2\pi^{2}n^2}{2m^{*}N^2a^{2}}$ at the $\Gamma$ point. As shown in Fig.\ref{Fig3}, at low $n$ and thick films $N$, the NFE model gives consistent results with the DFT calculations. However, at larger $n$ and for thin films, i.e., small values of $N$, the discrepancy between the NFE model and DFT calculations becomes apparent. This is expected since ,in bulk SVO, the NFE model gives parabolic energy dispersion $\frac{\hbar ^{2} k^{2}}{2m^*}$ which is only valid for a small momentum $k$. At larger $k$ the discrepancy between NFE (parabola) and TB model (cosine function) increases dramatically, as is shown in Fig.\ref{Fig1}(b) for the bulk.
For the same reason, the NFE model fails especially when the quantized energy is high (i.e., $n$ is large and $N$ is small), which explains the large difference in Fig.\ref{Fig3} between DFT and NFE model for such values of $n$ or $N$. In contrast, the energy dispersion of the TB model is in good agreement with DFT for small {\em and} large momentum $k$ , see Fig.\ref{Fig1}(b). We therefore expect a TB model to reliably describe the quantum well states.

\section{Tight binding (TB) Hamiltonian}
\subsection{First-principle based TB model for bulk SVO}
In this paper we take maximally localized Wannier orbitals for constructing a realistic TB Hamiltonian. The TB Hamiltonian has matrix elements
\begin{equation}
H_{\alpha\beta}(\vec{k})=\sum_{\vec{R}} t_{\alpha\beta}(\vec{R}) e^{\mathrm{i}\vec{k}\vec{R}} \; ,
\label{H}
\end{equation}
where $\vec{R}$ denotes lattice sites, $\alpha$ and $\beta$ denote orbitals in the Wannier basis, $t_{\alpha\beta}(\vec{R})$ represents a hopping integral from orbital $\alpha$ at site $0$ to orbital $\beta$ at site $\vec{R}$, and $\vec{k}$ is the wave vector. The Wannier projection on DFT calculated V $t_{2g}$ Bloch waves was performed with the Wien2Wannier package \cite{Kune20101888},
employing Wannier90 \cite{Mostofi2008685} for constructing
maximally localized Wannier orbitals.

For bulk SVO, we have a unit cell with a single V site and obtain three Wannier orbitals which are essentially $t_{2g}$ orbitals, but slightly hybridized with O$_{2p}$ orbitals \cite{zhong:epl12,Pavarini:njp05}. For simplicity we still denote these Wannier orbitals by $\alpha,\beta = xy, yz, xz$. All the orbitals are well localized with a localization function (variance) $\Omega$=1.89\AA$^2$ (defined in Ref.\ \onlinecite{Mostofi2008685}). For the following, we introduce the notation $\vec{R}=(l_x,l_y,l_z)=l_x \vec{e}_{x}+l_y \vec{e}_{y}+l_z \vec{e}_{z}$, where $\vec{e}_x$, $\vec{e}_y$, and $\vec{e}_z$ are lattice vectors along $x$, $y$ and $z$ direction, respectively, and $l_x$, $l_y$, and $l_z$ are integer numbers.

Through the Wannier projection, we obtain all hopping terms and construct a TB Hamiltonian according to Eq.(\ref{H}) which exactly reproduces the DFT calculated $t_{2g}$ bands as shown in Fig.\ref{Fig1}(b). All the major hopping terms are listed in Table \ref{bulk}. The $\vec{R}=(0,0,0)$ terms represent the local crystal field energies which is the same for the three orbitals due to cubic symmetry (often denoted as $\varepsilon$), and zero for inter-orbital elements such as $t_{xy \, yz}$. For $\vec{R}=(0,0,1)$, the inter-orbital hopping term is null due to symmetry, see positive (red) and negative (blue) lobes in Fig.\ref{Fig1}(a). The intra-orbital hopping term for $yz$ and $xz$ orbitals is large (-0.259eV) because these orbitals expand in the $z$ direction, while it is small (-0.026eV) for the $xy$ orbital which does not. 

Analyzing all hopping terms $t_{\alpha\beta}(\vec{R})$, we identify two basic characteristic features: (i) All the inter-orbital hopping terms are null or negligibly small, i.e., for $\vec{R}= (0,0,1)$ or $(0,0,0)$ they are
 exactly zero and for $\vec{R}=(1,1,0)$ and $(0,0,2)$ they are tiny (0.009eV or even less), see Table \ref{bulk}. As a result, the inter-orbital hopping process can be ignored to a very good approximation; all three orbitals are decoupled and can be treated separately. (ii) Along any specific direction, the next nearest neighbor hopping term (with $|l_z|$ $\geqq$ 2) is generally small. Hence, the nearest neighbor hopping already yields a good description for bulk SVO.
\begin{table}[b]
\begin{ruledtabular}
\caption[Tab1]{Hopping integral $t_{\alpha\beta}(\vec{R})$ in the maximally localized Wannier basis
for bulk SVO between orbital $\alpha$ at site 0 and orbital $\beta$ at site $\vec{R}$. $\vec{R}=(0,0,0)$ indicates the local energy term; $\vec{R}=(0,0,1)$ and $\vec{R}=(0,0,2)$ are 
 the nearest and next nearest neighbor along the $z$ direction, respectively. All values are in units of eV.}
\begin{tabular}{l l l l l l l}
 $t_{\alpha\beta}(\vec{R})$ & $\vec{R}$=(0,0,0) & (0,0,1) & (0,0,2) & (0,1,1) & \\
\hline
$xy$,$xy$ & 0.579 & -0.026 & 0.000 & 0.005 \\
$yz$,$yz$ & 0.579 & -0.259 & 0.007 & -0.082 \\
$xz$,$xz$ & 0.579 & -0.259 & 0.007& 0.005 \\
$xy$,$yz$ & 0 & 0 & 0.000 & 0.009 \\
\end{tabular}
\label{bulk}
\end{ruledtabular}
\end{table}

\subsection{Simplified TB model for bulk SVO}
Considering the two characteristic features of the hopping terms mentioned above, we can further simplify the TB model by decoupling the three orbitals and taking only the nearest neighbor hopping in Eq.(\ref{H}). For each orbital $\alpha$ along a given orientation such as $z$, the Hamiltonian in Eq.(\ref{H}) is then reduced to
\begin{equation}
H^{\alpha}(k_{z})=\varepsilon^{\alpha}+2t^{\alpha}\cos k_{z}a \; , 
\label{expression}
\end{equation}
 where $k_z$ is the wave vector along $z$, $\varepsilon^{\alpha}$ and $t^{\alpha}$ depend on $k_x$, $k_y$ and orbital $\alpha$. The more detailed expression and derivation can be found in the Appendix. 
Let us note here that $\varepsilon^{\alpha}$ and $t^{\alpha}$ can be either obtained from the $\vec{R}$=(0,0,0), (0,0,1) and (0,1,1) TB hopping elements,
or by a direct fit to the DFT band width and center of gravity in the given direction (here $z$). Both procedures yield similar results, see the Appendix.
We employ the latter in the following since this also mimics some of the effects of the other, smaller hopping elements. Along the $z$-direction this yields $t^{yz}=-0.475\,$eV, $\varepsilon^{yz}=-0.01\,$eV for the $yz$ (and $xz$) orbital; and $t^{xy}= -0.03\,$eV for the $xy$ orbital; $\varepsilon^{xy}=\varepsilon^{yz}+2t^{yz}-2t^{xy}$ preserves the degeneracy of $t_{2g}$ orbitals at $\Gamma$.


So far, we have simplified the TB Hamiltonian to a dispersion $2t\cos ka$ of nearest-neighbor-hopping type which allows us to treat all directions and orbitals independently.
 For small $k$, we now perform a Fourier expansion and obtain $\frac{\hbar ^{2} k^{2}}{2m^*}$ with $m^* =-\frac{\hbar ^{2}}{2a^2t}$. The obtained $m^*$ for $yz$ is 0.53$m_{e}$, which is very comparable to the NFE fitting value 0.56$m_{e}$. When $k$ is large however, $2t \cos ka$ gives a much better description than the NFE model, as shown in Fig.\ref{Fig1}(b). The physical origin for this discrepancy is the more localized nature of $d$ electrons in perovskite oxides materials. Considering the cosine energy-momentum dispersion and the negligible next nearest neighbor hopping term, we hence argue that electrons in SVO are quite tightly bound to the V atoms and their movement has the form of a hopping process from one site to a neighboring site. The NFE model of freely moving electrons is not applicable.

\begin{figure}[t!]
\includegraphics[width=\columnwidth]{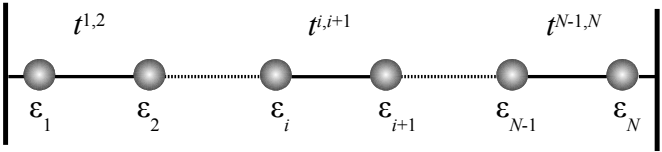}
\caption{Schematic figure of the effective one-dimension single-orbital TB model for describing electrons geometrically confined in ultrathin films. Here, $\varepsilon$ is the a local potential, and $t$ is the hopping term between nearest neighbors. The confinement is characterized by cutting the hopping term $t$ from the outmost sites ($i$=1, $N$) to vacuum.}
\label{Fig4}
\end{figure}
\subsection{TB model for SVO thin films}
For thin films, the unit cell contains $N$ Vanadium sites labeled by $i=1,2..N$. The corresponding 3$N$ Wannier orbitals are centered at each V site and have $t_{2g}$ orbital character. Following a similar procedure as in bulk SVO,\cite{zhong:epl12} a first-principles based TB Hamiltonian can be expressed in matrix form similar to Eq.(\ref{H}). Such a TB model can exactly reproduce the DFT results, as shown in Fig.\ref{Fig2} for $N$=6 layers.

\begin{table}[b]
\begin{ruledtabular}
\caption[Tab2]{Site and orbital dependent hopping integrals of SVO thin films with $N$=6. The first and second column is the on-site energy of $xy$ and $yz$ orbitals of each site $i$; the third and forth column the hopping integrals along the $z$ and $y$ direction for $yz$ orbitals, i.e., $t^{i-1,i}_{yz,yz}(0,0)$ and $t^{i,i}_{yz,yz}(0,1)$,respectively; the fifth column the hopping integrals along the $y$-direction for the $xy$ orbitals, i.e., $t^{i,i}_{xy,xy}(0,1)$. All values are in units of eV.}
\begin{tabular}{l l l l l l l}
& $yz$ & $xy$ & $yz$ along $z$ & $yz$ along $y$ & $xy$ along $y$ \\
\hline
1st V & 0.508 & 0.436 & 0 & -0.224 & -0.260\\
2nd V & 0.599 & 0.594 & -0.242 & -0.262 & -0.259\\
3rd V & 0.584 & 0.583 & -0.255 & -0.258 & -0.259\\
\end{tabular}
\label{interface}
\end{ruledtabular}
\end{table}

In contrast to bulk SVO, this thin film now has 18 Wannier orbitals which are centered around the 6 V sites and which have a similar character as bulk Wannier orbitals. Nevertheless, the Hamiltonian has some essential changes. One major change arises from the geometric confinement of thin films. In thin films, the lattice vector $\vec{R}$ becomes two dimensional with $(l_{x},l_y)=l_x \vec{e}_{x}+l_y \vec{e}_{y}$; the previous $l_{z}$ component now points to different V sites
{\em within} the unit cell and such a hopping is henceforth denoted by 
$t^{i,i+1}(0,0)$. In $k$ space this translates to a band structure which is dispersionless in the $z$ direction, but now we have $N$-times more bands.

Table \ref{interface} lists the calculated hopping integrals of the Wannier orbitals. Clearly there is no hopping from the surface layer (1st V layer)
to the vacuum, while there is a large hopping term ($-0.242$eV) between 1st and 2nd V layer. In contrast, all other layers contain hopping terms of similar magnitude to {\em two} neighboring sites along the $\pm z$ direction. In this sense, the predominant effect of the geometric confinement is to cut the hopping term from surface layer to vacuum. This simply reflects that electrons are not allowed to move outside the thin films, as illustrated in Fig.\ref{Fig4}. Such a geometric confinement plays a key role in quantum well states of SVO thin films.

There is a second important effect induced by the surface caused 
by the relaxation of the surface atoms: the surface Sr atom shifts inwards 0.12\AA \ and the surface O atom outwards 0.06\AA, due to surface dangling bonds. This changes the local crystal fields in the surface layer and to a lesser extend in the neighboring subsurface layers.  As listed in the Table \ref{interface}, the local crystal field energies (1st and 2nd column) become site and orbital dependent. The biggest effect is observed for the
$\varepsilon^{xy}$ of the surface layer which has a 0.16eV lower energy than in the second layer.  This local potential is responsible for the
 DFT pronounced level splitting of the $xy$ orbitals at $\Gamma$, see Fig.\ref{Fig2}. We note $\varepsilon$ and $t$ converge to the bulk values very quickly; already for the third layer the difference to the bulk value is small. In this sense, a surface potential well will be formed. In the following Sections, we will show that such a potential well plays a crucial rule for the surface confinement of 2D electron gas at STO surfaces and LAO/STO interfaces.
Hence, we need to include this effect for surfaces and  interfaces (Section 
\ref{Sec:surface}), whereas it is of lesser relevance and hence has not been taken into account for the thin film geometry (Section 
\ref{Sec:thinfilm}).

\subsection{Simplified TB model for confinement in SVO thin films}
\label{Sec:thinfilm}
To obtain an intuitive physical picture, we will again simplify the first-principles based TB model. We first ignore the surface effect (surface or interface potential well), and focus on the geometrical confinement of the hopping term only. We here employ the same approximation and parameters as in the simplified TB model for bulk in Eq.(\ref{expression}). That is, for a given orbital and specific $k_{x}$, $k_y$, we have a one-dimensional intra-band TB hopping.
For the thin layer, this single-band TB hopping is confined within $N$ sites. Hence we simply cut the hopping term from surface layer to vacuum, as illustrated in Fig.\ref{Fig4} and justified by 
Table \ref{interface}. The Hamiltonian is then expressed as a $N\times N$ matrix
\begin{equation}
\left(\begin{array}{cccccc}
\varepsilon & t & 0 & 0 & 0 & 0\\
t & \varepsilon & t & 0 & 0 & 0\\
0 & t & \varepsilon & t & 0 & 0\\
0 & 0 & ... & ... & t & 0\\
0 & 0 & 0 & t & \varepsilon & t\\
0 & 0 & 0 & 0 & t & \varepsilon\end{array}\right) \; ,
\label{matrix}
\end{equation}
Here $t$ and $\varepsilon$ depend on $k_{x}$, $k_y$ and $\alpha$ in the same way as in Eq.\ (\ref{expression}). The eigenvalues of the matrix are the quantized energies of the quantum well states that are confined to the thin film.  For such a tri-diagonal matrix the eigenvalues have a simple analytical expression:
\begin{equation} 
\varepsilon+2t\cos (\frac{\pi n }{N+1}) ; n=1, 2, ...,N \; ,
\label{quantized}
\end{equation}
where the quantum number $n$ indexes the $N$ quantized energy levels 
emerging from the confinement in the $z$ direction.
At $\Gamma$, we take the bulk values $t$=-0.475eV and $\varepsilon$=-0.01eV for the $yz$ orbital. 
The quantized energies of Eq.(\ref{quantized}) give much better results 
than the quantized levels $\frac{{\hbar}^2\pi^{2}n^2}{2m^{*}N^2a^{2}}$ of the NFE model, as the comparison with DFT in Fig.\ref{Fig3} shows. For a low quantum number $n$ and a thick film with large $N$, the quantized energies are small, and the two models give consistent results. However, for larger quantum number $n$ or thin films with small $N$, the TB gives much better results. This is expected, since the TB model yields a good description for both a small and large momentum $k$, as is shown for bulk SVO shown in Fig.\ref{Fig1}(b). 
While the DFT clearly shows the superiority of the TB model,
experimentally more data are needed for a clear statement 
in this respect. This is possible by growing thinner films (small $N$),
where the separation between NFE and TB model becomes apparent.

 Next, we will consider a surface potential well as a further source of confinement. In principle this can be done for the SVO thin layer. However, in the
case of the $yz$ orbitals of Fig.\ \ref{Fig3} the quantum well state spreads over all layers of the thin
film, so that the surface potential hardly affects the results of Fig. \ref{Fig3}. This is different for the $xy$ orbitals, as here the wave functions are localized within single layers and the surface layer has a rather different potential (see Table \ref{interface}).
The effects are however even more dramatic for a (single) surface or interface. Here, without
 surface/interface potential the wave functions spread throughout the (semi)infinite structure, and the behavior is hence the same as in the bulk.
In this situation, the surface potential is needed to generate a quantum well state,
and we hence study STO surfaces and interface in the following Section.

\section{STO surfaces and LAO/STO interfaces}
\label{Sec:surface}
\subsection{TB model}
In contrast to SVO ultrathin films, where electrons are geometrically confined within the thin films by cutting the hopping terms from the two surfaces into vacuum, STO surfaces (or LAO/STO interfaces) is a semi-infinite system with only one surface (or interface) where the hopping term is cut. 
Hence the cut hopping in itself is not sufficient for
a quantum confinement and quantized energy subbands.
An attractive potential at the surfaces is required to trap electrons
in a 2D conducting sheet. 

\begin{figure}[t!]
\includegraphics[width=\columnwidth]{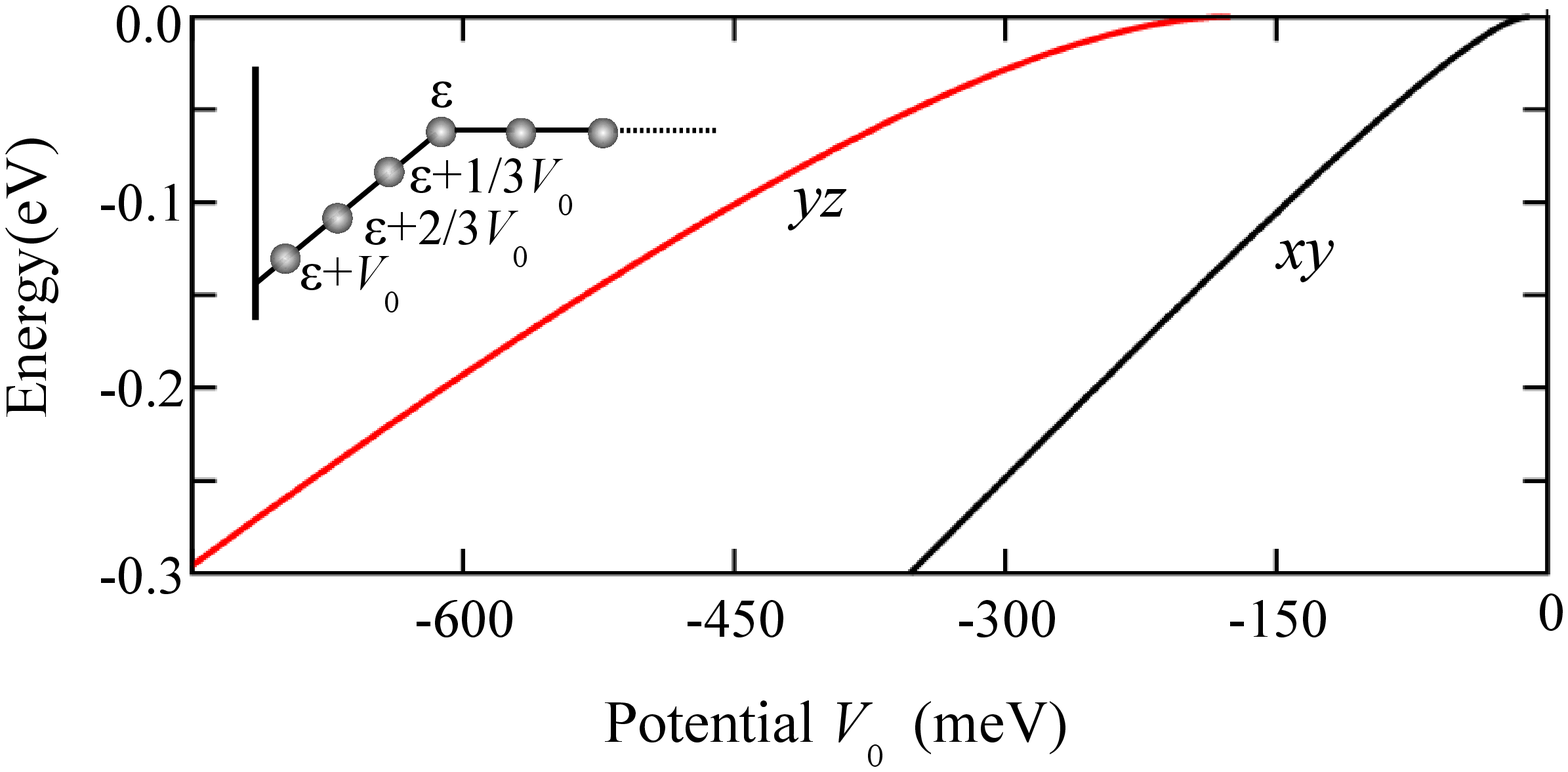}
\caption{Energy gain of the lowest $xy$ (black) and $yz$ (red) energy eigenstate due to a surface potential of strength $V_{0}$ and a width of three layers as depicted  in the inset.}
\label{Fig5}
\end{figure}

Generally, the surface potential can be generated by two sources: extrinsic defects such as accumulation of defects at the surface, and intrinsic surface effects such as atomic relaxation. To calculate the former one, we need the distribution of the defects, and then solve the potential well and 2DEG self-consistently. In this case, the quantitative strength of the extrinsically induced surface potential depends on experimental details and might vary considerably. This extrinsic surface potential is not considered in our work and would add to the latter intrinsic one which can be well included by DFT calculations.
Instead, we consider the intrinsic surface potential of a defect-free surface 
due to the atomic relaxation at the surface.

Indeed, both the DFT calculation of the latter  \cite{Popovic:prl08,Janicka:prl09,zhong:epl12,Delugas:prl11} and experiment \cite{Santander:nat11, Yoshimatsu:prl08, Mannhart:sci10,Slooten:prb13} show  a potential well of width 3 to 4 layers and of depth  0.2$\sim$0.3eV at the STO surface. The DFT calculated band structure is very similar to the case of SVO thin films, which indicates some general behavior of perovskite oxide heterostructure, such as the splitting between $xy$ and $yz$ bands, quantized $yz$ subbands, and that the lowest $yz$ orbital has a large spread into the bulk layers.

\begin{figure}[t!]
\includegraphics[width=\columnwidth]{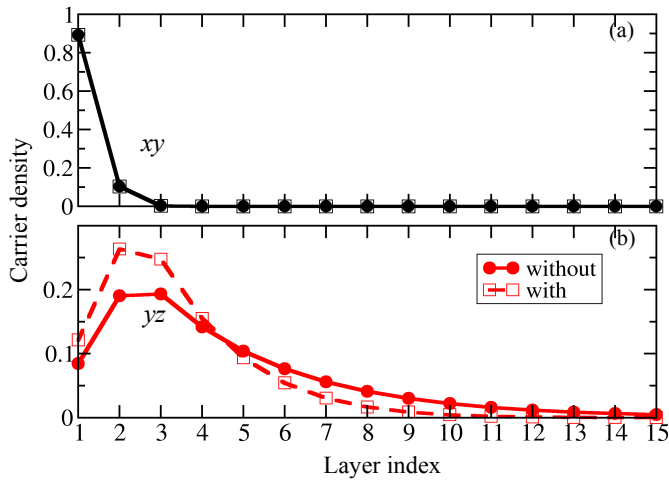}
\caption{Layer-resolved charge distribution of the lowest quantized state of $xy$ (a) and $yz$ (b) character when confined by a surface potential $V_{0}=-240$meV as depicted in the inset of Fig.\ref{Fig5}. 
we show the charge distribution without (filled circles) and with
 an external electric field of -5mV per unit cell lattice vector 
(unfilled squares). The layer with index 1 denotes the surface layer.}
\label{Fig6}
\end{figure}

To model the surface potential well, 
we introduce a site dependent $\varepsilon$, as depicted in the inset of Fig.\ref{Fig5}. The width of the surface potential well is taken to be three unit cell as suggested by both, DFT and experiment. 
If we assume that 
the $xy$ and $yz$ orbitals have the same local potential $\varepsilon$, i.e, the same $V_0$ in Fig.\ref{Fig5}, the two orbitals only differ regarding the magnitude of their hopping terms: $t^{xy}=$-0.03eV and $t^{yz}=$-0.475eV for hopping along the $z$ direction. We cut the hopping term from the surface layer to vacuum, and increase the thickness $N$ up to 100 sites to simulate the semi-infinite condition until the quantized energies are converged.

In the TB model we have to calculate the eigenvalues of the matrix Eq.(\ref{matrix}) supplemented by a layer dependent $\varepsilon$. The quantized $xy$ and $yz$ energies which we obtained numerically are plotted in Fig.\ref{Fig5} as a function of the strength $V_0$ (see the inset of Fig.\ref{Fig5} for the relation to $\varepsilon$; the bulk reference energy is set to zero). Electrons are confined in a quantum well surface state if 
and only if the lowest energy (relative to the bulk energy) is negative in Fig.\ \ref{Fig5}. Otherwise electrons are not confined at the surface, but become 3D bulk like. For $V_0>$-30meV, no 2D electron gas can be formed at the surface; both $xy$ and $yz$ electrons will spread into the bulk layers. For -30meV$ >V_0> $-200meV, only $xy$ electrons are 2D confined, whereas electrons in the $yz$ orbitals still spread into the bulk. This is because the $t^{yz}$ hopping is much larger than $t^{xy}$, and hence $yz$ orbitals extend more easily into the bulk. Eventually for a potential strength $V_0<$-200meV, both $xy$ and $yz$ electrons can be confined at the surface like and a 2DEG is formed.

Both the DFT \cite{Janicka:prl09,zhong:epl12} calculated and the experimental \cite{Yoshimatsu:prl08} observed surface potential $V_{0}$ is about -300 to -200meV per three unit cell. Hence, we conclude that $xy$ carriers are always localized at the surface, whereas $yz$ carriers are on the verge of a 2D confinement. Our results therefore suggest that whether $yz$ states are quantized or not is very sensitive to surface details. This might explain why Santander-Syro {\it et al.}\cite{Santander:nat11} observed a $yz$ subband at STO surfaces, whereas Meevasana {\it et al.}\cite{Meevasana:natm11} did not. 

If we take $V_{0}$=-240meV, the splitting between $xy$ and $yz$ states at $\Gamma$ will be about 150meV, which is compatible with experimental and DFT values.
 For this typical surface potential strength, we plot the charge distribution of the lowest quantized $xy$ and $yz$ states in Fig.\ref{Fig6}.
 The $xy$ state is strongly localized at the surface layer (upper panel of Fig.\ref{Fig6}), and hence its quantized energy in Fig.\ref{Fig5} also basically reflects the local surface potential $V_{0}$. In a similar way, the second quantized state is localized at the sub-surface layer, and its energy reflects the local potential of sub-surface site, i.e., $2/3\, V_{0}$ (not shown). On this basis, we argue that the energy of the $xy$ subbands \cite{Santander:nat11, Meevasana:natm11} can serve as a measure of the surface potential well. In contrast, even the lowest $yz$ state has a very long tail extending $\sim$10 unit cells into the bulk \cite{Stengel:prl11,Khalsa:prb12,Park:arxiv13,Plumb:arxiv13}. That is, even though the surface layer has the lowest local potential, the lowest quantized $yz$ state has actually a large contribution from the second and third layer. We emphasize that this TB result is consistent with DFT \cite{Popovic:prl08,Janicka:prl09,zhong:epl12,Delugas:prl11}.

\subsection{Applying an external electric field}
Since the lowest $yz$ subband is on the verge of a 2D confinement for a realistic surface potential well, an external electric field might strongly influence its 2D properties. We hence apply an external electric field which 
together with the induced polarization yields an effective internal field
which we assume to be -5mV per unit cell. Hence, we add a potential of -5meV per layer. Considering the huge polarization of STO, such an internal electric field is experimentally feasible \cite{Caviglia:prl10b}. Fig.\ref{Fig6} shows that the charge distribution of the lowest $yz$ state changes dramatically, whereas the $xy$ orbital is virtually unaffected. This striking result indicates that applying an electric field cannot tune the $xy$ charge carries, but does tune the $yz$ charge carriers. This result hence indicates that electric field tunable properties such as superconductivity \cite{Caviglia:nat08}, spin-orbit coupling\cite{Caviglia:prl10b,Shalom:prl10b}, and mobility \cite{Bell:prl09,Kim:prb10} stem predominantly from $yz$ charge carriers. The fact that the lowest $yz$ subband is on the verge of the 2D confinement might be the key for understanding all the puzzling behavior at LAO/STO or STO surfaces.

\section{Discussion and conclusion}
In this paper, we developed first-principles based tight-binding (TB) models with hopping $t$ and site dependent potential $\varepsilon$ to study the quantum condiment of perovskite oxide heterostructures of two specific cases: (i) SVO ultrathin films, where electrons are geometrically confined by cutting a hopping term $t$ from surface to vacuum; (ii) STO surfaces or LAO/STO interfaces, where electrons are confined by a surface potential well as described by a layer-dependent potential $\varepsilon$. In both cases, we have shown that a simple TB model gives a much better and more reliable description of $d$ electrons in transition metal oxides than a nearly free electron (NFE) model.

Already the two hopping parameters in the two inequivalent nearest neighbor directions of the $t_{2g}$ orbitals, describes the complex DFT and experimental bandstructure of  SVO films well, including the
 orbital-selective quantum well states. By means of the TB model, we find that the discrete energy levels at $2t \cos(\frac{\pi n }{N+1})$ with quantum number $n=1 \ldots N$, in contrast the NFE model with levels at $\frac{{\hbar}^2\pi^{2}n^2}{2m^{*}N^2a^{2}}$. For STO surfaces and LAO/STO interfaces with a reasonable surface potential well, $xy$ states are always localized as 2D carriers which is
directly reflected by the discrete $xy$ energy levels. In contrast, the lowest $yz$ state is on the verge of 2D confinement and has a much large extension into the bulk layers. As a consequence we show that the $yz$ charge distribution, 
but not the $xy$, can be tuned by an experimental accessible electric field.

Generally speaking, the TB model can give an intuitive physical picture as simple as the NFE model but is much more accurate for surfaces, interfaces and superlattices of transition metal oxides. All parameters of the TB model can be determined from DFT through a Wannier function projection. The TB model hence provides the basis for subsequent calculations such as large-scale simulation, transport properties or many-body effects. 
Incorporating also magnetism, correlations, spin-orbit coupling, and superconductivity will complete the TB model and allow us to figure out all the essential physics of oxide heterostructures.

\section*{Acknowledgment}
ZZ acknowledges financial support by the Austrian Science Fund through
the SFB ViCoM F4103,  QFZ by NSFC (11204265), the NSF of Jiangsu Province (BK2012248), and KH by the
 European Research Council under the European Union's Seventh 
Framework Programme (FP/2007-2013)/ERC
 through grant agreement n.\ 306447. Calculations have been done on the Vienna Scientific Cluster~(VSC).\\


\section{Appendix}
Let us consider the energy-momentum dispersion along the $z$ direction for fixed $k_x$,$k_y$:
\begin{eqnarray*}
H_{\alpha\beta}(k_x,k_y) (k_z) & = & \sum_{l_x, l_y,l_z} t_{\alpha\beta}(l_x,l_y,l_z) e^{\mathrm{i}(l_xk_x+l_yk_y+l_zk_z)} \\
 & = & \sum_{l_x, l_y} \sum_{l_z} t_{\alpha\beta}(l_x,l_y,l_z) e^{\mathrm{i}(l_xk_x+l_yk_y)} e^{\mathrm{i}l_zk_z}
\end{eqnarray*}
Since the next nearest neighbor hopping term $\vec{R}$=(0,0,2) as listed in Table \ref{bulk} is negligible, we consider only the nearest neighbor hopping along $z$ direction with $|l_z|\leqq 1$. Due to the inversion symmetry of bulk SVO, $t_{\alpha\beta}(l_x,l_y,-1)=t_{\alpha\beta}(l_x,l_y,1)$, 
\begin{eqnarray*} 
H_{\alpha\beta}(k_x,k_y) (k_z) &=& \sum_{l_x, l_y,0}t_{\alpha\beta}(l_x,l_y,0)e^{\mathrm{i}(l_xk_x+l_yk_y)} + \\
&& \sum_{l_x, l_y}t_{\alpha\beta}(l_x,l_y,1) e^{\mathrm{i} (l_xk_x+l_yk_y)} 2\cos(k_za).
\end{eqnarray*}
Considering furthermore that the inter-orbital hopping term is negligible, we the three orbitals decouple with an intra-orbital Hamiltonian
\begin{equation}
H_{\alpha\alpha}(k_z)=\varepsilon^{\alpha}+2t^{\alpha}\cos(k_za).
\label{final}
\end{equation}
Here, $\varepsilon^{\alpha}=\sum_{l_x, l_y}t_{\alpha\beta}(l_x,l_y,0)e^{\mathrm{i}(l_xk_x+l_yk_y)}$ and $t^{\alpha}=\sum_{l_x, l_y}t_{\alpha\beta}(l_x,l_y,1) e^{\mathrm{i}(l_xk_x+l_yk_y)}$. The simple analytical from of Eq.\ (\ref{final}) accounts for the most important hopping terms of $t_{2g}$ orbitals, and 
allows us to easily compare
 the energy-momentum dispersion to the theoretical, e.g., DFT, bandstructure
or ARPES experiments. We still need to determine $\varepsilon^{\alpha}$ and
$t^{\alpha}$, which depend on the orbital and direction considered.

For instance, in case of the $yz$ orbital and $z$-direction the hopping terms are as listed in Table \ref{bulk}: $t(0,0,0)\equiv t_0=0.579\,$eV, $t(0,0,1)\equiv t_1=-0.259\,$eV, $t(1,0,0)\equiv t_2=-0.026\,$eV, $t(0,1,1)\equiv t_3 =-0.082\,$eV; for the $xy$, $xz$ orbital related terms have to be taken. From these, we obtain the effective parameters $\varepsilon$ and $t$ for three orbitals at fixed $k_x$, $k_y$ for the dispersion along the $z$ direction:
\begin{eqnarray*}
\varepsilon^{xy} & = &t_{0}+2t_{1}\cos k_{x}a+2t_{1}\cos k_{y}a+4t_{3}\cos k_x\cos k_y \\
\varepsilon^{yz} & = &t_{0}+2t_{2}\cos k_{x}a+2t_{1}\cos k_{y}a \\
\varepsilon^{xz} & = &t_{0}+2t_{1}\cos k_{x}a+2t_{2}\cos k_{y}a \\
t^{xy} & = &t_{2} \\
t^{yz} & = &t_{1}+2t_{3}\cos k_{y}a \\
t^{xz} & = &t_{1}+2t_{3}\cos k_{x}a 
\end{eqnarray*}
If we focus on the energy dispersion from $\Gamma=(0,0,0)$ to $(0,0, \pi/a)$, we set $k_x$=0, $k_y$=0. For the $yz$ orbital, we then obtain 
$t^{yz}=t_{1}+2t_{3}=$-0.423eV and $\varepsilon^{yz}=t_{0}+2t_{1}+2t_{2}=0.009$eV.
In this direction, the $xz$ orbital has the same parameters due
to cubic symmetry. 
For $xy$ orbital on the other hand, the two effective TB parameters are $t^{xy}=-0.026$eV, and $\varepsilon^{xy}=t_{0}+4t_{1}+4t_{3}=-0.785$eV.
At $\Gamma$, all three orbitals are degenerate and have the energy $t_{0}+4t_{1}+2t_{2}+4t_{3}$.

Alternatively, we can fit $t^{yz}$ and $\varepsilon^{yz}$ directly to the DFT bandstructure: The band dispersion of the $yz$ orbital from $\Gamma$(0,0,0) to Z$(0,0,\pi/a)$ is 1.90eV. Hence, 
 Eq.(\ref{expression}) and DFT give the same band width for $t^{yz}=-1.90\,$eV$/4=-0.475\,$eV. The center of gravity of the band allows us to determine
$\varepsilon^{yz}= -0.01\,$eV. This fit well agrees with the above parameters determined from the TB hopping parameters. The same is true for the $xy$ orbital.
Here, the DFT band width is 0.12eV, and hence $t^{xy}=-0.03\,$eV; $\varepsilon^{xy}=\varepsilon^{yz}+2t^{yz}-2t^{xy}$ preserves the degeneracy of the $t_{2g}$ orbitals at $\Gamma$. 


\begin{thebibliography}{45}
\bibitem[{Foo()}]{Footnoot1}
\bibinfo{journal}{Spin orbit coupling will mix $t_{2g}$ characters near
  $\Gamma$ and significantly modify the effective masses.}  (????).
\end{thebibliography}

\end{document}